\documentclass{amsart}
\usepackage{graphicx}
\newtheorem{thm}{Theorem}[section]

\newtheorem{prop}[thm]{Proposition}
\theoremstyle{definition}
\newtheorem{defn}[thm]{Definition}
 \theoremstyle{example}
\newtheorem{exmp}[thm]{Example}
\numberwithin{equation}{section}
\theoremstyle{remark}
\newtheorem{remk}[thm]{\bf Remark}
\numberwithin{equation}{section}
\newenvironment{prf}{\noindent{\bf Proof}}{\\ \hspace*{\fill}$\Box$ \par  }
\begin{document}

\title[A Digital version of Green's Theorem and its application to the coverage problem in Formal Verification]{A Digital version of Green's Theorem and its application to the coverage problem in Formal Verification}%
\author{Eli Appleboim$\dag$ \& Emil Saucan$\ddag$}%
\address{Technion, Department of Electrical Engineering$\dag$ \& Technion, Department of Mathematics and Ort Braude, Software Engineering Department$\ddag$}%
\email{eliap@ee.technion.ac.il$\dag$ \& semil@tx.technion.ac.il$\ddag$}%

%\thanks{}%
%\subjclass{}%
%\keywords{}%

%\date{}%
%\dedicatory{}%
%\commby{}%
% ----------------------------------------------------------------
\begin{abstract}
 We present a novel scheme to the coverage problem, introducing a quantitative
 way to estimate the interaction between a block an its environment. This is achieved by setting a
 discrete version of Green's Theorem, specially adapted for Model Checking based verification
 of integrated circuits.
\\ This method is best suited for the coverage problem since it enables one to quantify the incompleteness or, on
the other hand, the redundancy of a set of rules, describing the model under verification.
 Moreover this can be done continuously throughout the verification process, thus enabling  the user to pinpoint the stages at which incompleteness/redundancy occurs.
\\ Although the method is presented locally  on a small hardware example, we additionally show its possibility to provide precise
coverage estimation also for large scale systems. We compare this method to others by checking it on the same
test-cases.
\end{abstract}
\maketitle
% ----------------------------------------------------------------
\section{Introduction}
While automatic verification methods (s.a. Model Checking, etc.) permit quite accurate formulation of rules
describing concurrent systems, the coverage problem is still open. This problem can be stated
 as ones ability to know that a certain set of rules covers all possible behaviors of the
system and if so, whether this set is optimal, in the sense that it does not contain redundancies.
\\ This problem, besides being a very challenging research problem, also
plays a crucial role in industrial implementation of verification methods, in aspects of manpower, time and, (of
course, finance.
\\ While different methods to attack this problem where proposed (see \cite{kgg}, \cite{hkhz}, \cite{as}), it is still largely open.
\\ In this paper we present a novel scheme to the {\em coverage problem}, introducing a quantitative
 way to estimate the interaction between a block an its environment. This is achieved by setting a
 discrete version of Green's Theorem, specially adapted for Model Checking based verification.
 \\This work was inspired  by the well known principle of Model Checking that a well written environment dictates the
 formulation of the system's rules; indeed the rules governing the system and the description
 of its physical proprieties should be  regarded as almost mirror images of each other.
 On a more theoretical level,
 resides the idea of viewing a flow of information in an analogous way to
 the electromagnetic (or energy) flow and the computation of its mean flux, i.e. pressure.  The main idea behind  the results
 presented herein is the adaptation of the symmetry principle to the flow setting, adaptation which
 states that the mean pressure of information is constant, (i.e. the informational system is in dynamic
 equilibrium).
\\ This method is best suited for the coverage problem since it enables one to quantify the incompleteness or, on
the other hand, the redundancy of a set of rules, describing the model under verification.
 Moreover this can be done continuously throughout the verification process, thus enabling  the user to pinpoint the stages at which incompleteness/redundancy
 occurs. The method we present here does not permit the complete automation of the coverage-checking process (thus
 making  the verifier's role redundant), since it doesn't guarantee completeness of
 coverage, but only that inconsistencies or redundancies are discovered. Thus it is yet another instrument in the arsenal of the experienced verifier, and one that is extremely easily to use without any further specialization. Moreover, it can be readily added as supplementary feature, to any existing industrial machine.
 \\ The paper is organized as follows: In Section 2 we give the basic preliminaries and show how to
formulate Green's Theorem in the context of pressure of information. In this section this is done in a basic,
local setting of individual blocks  composing a system (i.e. silicone "chip"). In Section 3 we compare this method
to the one presented in \cite{kgg}, by applying it to the same test-cases presented therein. In Section 4 we show
how this method naturally extends globally to large scale units and can be implemented up to the level of
integrated circuits. This extending  ability enables one to get the most out of this method, since it makes it
possible to unify all stages of the development, from the architect, through the designer, to the verifier.
Moreover, this method relieves the verifier from the need of unnecessary presumption that the verification
 of a neighboring unit has been done correctly; instead it gives verifying tools to
easily quantify this correctness. Finally, in Section 5 we gives outlines for future research.
\section{Theoretical Setting}
\subsection{Definition and Notations}
In this section we give the basic background and notations. A brief discussion on Green's Theorem is given in the
Appendix.
\begin{defn} A {\em block} $B$ is a punctured topological disk $B \simeq \bar{D} \setminus \{p_1,...,p_s\}, \newline s \geq 0$; where
$\bar{D}$ denotes the closed unit disk $\bar{D} = \{z\in C^2\,|\, |z| \leq 1\}$.
\\ The {\em environment} of a block $B$, $env(B)$ is the complement of $B \cup \{p_1,...,p_s\} =
\overline{R^2 \setminus D}$.
\end{defn}
A block and its environment share a common boundary and communicate with each other via {\em input/output}
 messages.
\begin{defn} An {\em information unit} is a {\em signal} $s$ that may have the values $0,1$.
\\ A {\em message} is a pair $m_k = (\pm \,\mathbf{n}; s_1,...,s_k)$, where $+\,\mathbf{n}$ is the unit normal
vector pointing outward form the block, toward $env(B)$, and $\{s_1,...,s_k\}$ is a set of information units.
\\ An {\em input} is a message of the form: $i = (-\,n;s_1,...,s_k)$;  while an {\em output} is a message of the form: $o =
(+\,n;s_1,...,s_k)$.
\end{defn}
The punctures also connect with the block via sending/recieving messages where the directions are $\pm
\,\mathbf{n}$ with respect to the boundary of a small disk neighborhood of a puncture point. A puncture will be
called a {\em sink} if all its messages are {\em outputs}, and a {\em source} if all its messages are {\em inputs}
(where the orientations of the normal vectors are considered with respect to the block $B$).
\begin{remk}
We refer mainly to typical punctures of sink or source types, but of course, "mixed" punctures  are also possible
(therefore they are to be considered).
\end{remk}
\begin{exmp} \label{eq:EXMP}The block illustrated in Fig.1 has -- relatively) to the outer
boundary -- an input message: ${\tt(-\,{\bf n}\,;ack)}$ and output message ${\tt(+\,{\bf n}\,;req,wr)}$. It also
has a puncture ${\tt p = s_{-}}$ of type source; its messages being: ${\tt (-\,{\bf n}\,; sign1)}$ and ${\tt
(-\,{\bf n}\,; sign2)}$.
\\ It should be noted that a message is appears only if its signals are {\em asserted},
i.e. as unit signals; that is we identify ${\tt(+\,{\bf n}\,;req,wr)}$ with the vector ${\tt(+\,{\bf n}\,; 1,1)}$
(or ${\tt(+ \,1\,; 1,1)}$). %\vspace*{-2cm}{\mbox{$env(B)$}}
\begin{figure}[h]
\begin{center}
\includegraphics[scale=0.3]{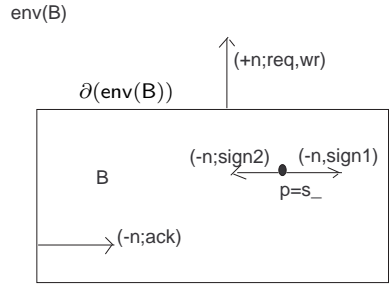}
\end{center}
\caption{A typical block}{\vspace*{-3.9cm}\hspace*{-2cm}\footnotesize$\sf{\partial(env(B))}$}
\end{figure}%\marginpar{\vspace*{-2cm}\hspace*{-7cm}\tiny Hey, Latex!! $\biguplus$}
\end{exmp}\vspace*{4cm}
\subsection{Main Theorem} %\marginpar{\footnotesize Good Title? $\rightarrow$}
 \begin{defn} For each  message  (i.e. input, output, sink/source) we define a {\em
 measure} $\mu$ -- the {\em information pressure} -- according to:
\[\begin{array}{l}
\mu(i) = \mu(\mathbf{n}; s_1,...,s_k) = k;\\
\mu(o) = \mu(-\,\mathbf{n}; s_1,...,s_l) = -l;\\
\mu(s_i) = \mu(\mathbf{n}; s_1,...,s_p) = p\, ;\\
\mu(s_o) = \mu(-\,\mathbf{n};s_1,...,s_q) = -q\,.
\end{array}\]
\end{defn}
\begin{remk}
Again, we allow the existence of "mixed" punctures. In this case the measure associated which such a puncture is
the arithmetic sum of its signals, considered with sign "$+$" if they relate to a output, and with a "$-$" if they
correspond to an input.
\end{remk}
\begin{exmp}
Consider the puncture $p$ with messages: $(+\,{\bf n}\,; s_1 ,s_2), \;(+\,{\bf n}\,; s_3)\,, \;\\ (+\,{\bf n}\,;
s_4 ,s_5)$. Then: $\mu(p) = 2 + 1 - 2 = 1$.
\end{exmp}
 With these notations we are in a position to formulate  Green's theorem for blocks:
\begin{thm} \label{thm:THM}  Let $B$ be a block of(with) outputs $o(B)$, inputs $i(B)$ and sink/sources $p(B) = p_{+}(B) +
p_{-}(B)$, where $p(B)$ denotes the set of punctures of the block $B$. Then the following holds:
\begin{equation}
\sum_{i \in i(B)}\mu(i) +  \sum_{o \in o(B)}\mu(o) +  \sum_{p \in p(B)}\mu(s) = 0 \label{eq:THM}
\end{equation}
\end{thm}
\begin{prf}
  Given a block $B$, every input/output signal is uniquely identified with a point on the boundary $Fr(B)$ with a
  length element given by the measure $\mu$ of the signal. Since overall-time information is conserved, applying Grenn's Formula on such a block gives:
\begin{equation}
 0 = \oint_{\partial(B)}\hspace{-0.4cm}{\footnotesize information}\, d\mu = \sum_{i \in i(B)}\mu(i) +  \sum_{o \in o(B)}\mu(o) +  \sum_{p \in p(B)}\mu(s)
\end{equation}
\end{prf}
\begin{exmp}
The following simple rule relative to the block of Example \ref{eq:EXMP} illustrates the method of implementing
Formula ~\ref{eq:THM}:
 \[{\bf AG}\,({\tt (\neg ack \wedge req \wedge wr)}\rightarrow {\bf AF}\,{\tt ack}) \qquad \rho\]
 Since ${\tt req}$ and ${\tt wr}$ are outputs, they both contribute with a "$+1$", while ${\tt ack}$, being an
input, adds a "$-1$" to the general balance, so the measure {\em variance} associated to the rule $\varrho$ is
$\Delta(\varrho) = 1 + 1 - 1 = 1$. Note that - as stated before - we count only the asserted signals.
\\ We shall show in {\em Section 3} how to add the variances of  individual rules in order to get the {\em global
 variance} $\Delta(B)$.
\end{exmp}
 \hspace*{-0.5cm} {\bf Note} \, Formula ~\ref{eq:THM} should be understood as a qualitative
indicator for the  coverage of a given set of rules, hence this set is assumed to satisfy the following
postulates:
\begin{enumerate}
\item Every input/output appears {\em at least} once in the set of rules, otherwise one could have, for instance the following set of rules, which consists
of only one formula: \[ \mathcal{S}_1\!: \;{\bf AG}\,({\tt req} \rightarrow {\bf AX }\,{\tt ack}) \qquad \rho_1\]
which satisfies ~\ref{eq:THM}, but evidently will not represent a complete set of rules for a realistic arbiter.
\item If a set of rules does not satisfy ~\ref{eq:THM}, it is guaranteed to be either incomplete or to have
redundancies. On the other hand, sets of rules may satisfy ~\ref{eq:THM}, but still be inconsistent, such as the
following:
\[ \mathcal{S}_2\!: \; \left\{ \begin{array}{ll}{\bf AG}\,({\tt req} \rightarrow {\bf AX }\,{\tt ack}) & \qquad \rho_1\\
                         {\bf AG}\,({\tt req} \rightarrow {\bf AX\, AX}\,{\tt ack}) & \qquad \rho_2
         \end{array} \right. \]
Indeed, if "{\tt ack}" is - as it normally does - a "pulse" signal, then $\rho_1$ and $\rho_2$ will not be always
satisfied  in tandem by any normal system.
\\Also, the following system contains a redundant rule:
\[ \mathcal{S}_3\!: \; \left\{ \begin{array}{ll}{\bf AG}\,({\tt req} \rightarrow {\bf AX }\,{\tt ack}) & \qquad \rho_1\\
                                               {\bf AG}\,({\tt req} \rightarrow {\bf AF}\,{\tt ack}) & \qquad \rho_3
         \end{array} \right. \]  since if $\rho_1$ holds, then $\varrho_3$ is obviously redundant.
However it still obviously satisfies ~\ref{eq:THM}, so a direct application of Theorem ~\ref{thm:THM} will not
reveal this fact.
\\ Therefore, it is imperative that the formalist satisfies the following:
\\ \\ \hspace*{-0.8cm}{\em Fairness Assumption}\ \ The set of rules consists only of relevant
 rules and does \hspace*{2.75cm}not contain deliberate redundancies.
\end{enumerate}
\newpage
\begin{remk} Although, as shown in the previous Note, the given method does not give a
completely automatic tool to solve the {\em coverage problem}, it gives the user, especially the skilled verifier,
a numerical assessment of the block's complexity. Two important conclusions ensue from this:
\\ On one hand, a well formulated set of rules enables one to actually compute the complexity of the internal
structure of the block, as this is expressed  by the pressures contributed by the punctures, thus allowing a
paradigm shift from the {\em black-box} concept to that of {\em semi-transparent} blocks.
\\ On the other hand, it gives the verifier of neighboring blocks a computational tool for checking
relative correctness along the common interface. This advantage becomes even more effective in pipelining units,
for which the boundary interface is the simplest possible.
\end{remk}
\section{Case Studies}
This work was partially motivated by the work of Katz, Grumberg and Geist (see \cite{kgg}). We will demonstrate
the application of ~\ref{eq:THM}  to the examples given in the paper mentioned above, and compare the results
obtained and the efficiency of both methods.
\\ Their main example consists of a synchronous arbiter $A$ having two
inputs: ${\tt req1}$ and ${\tt req2}$ and two outputs: ${\tt ack1}$ and ${\tt ack2}$ (see Fig. 2).
\begin{figure}[h]
\begin{center}
\includegraphics[scale=0.3]{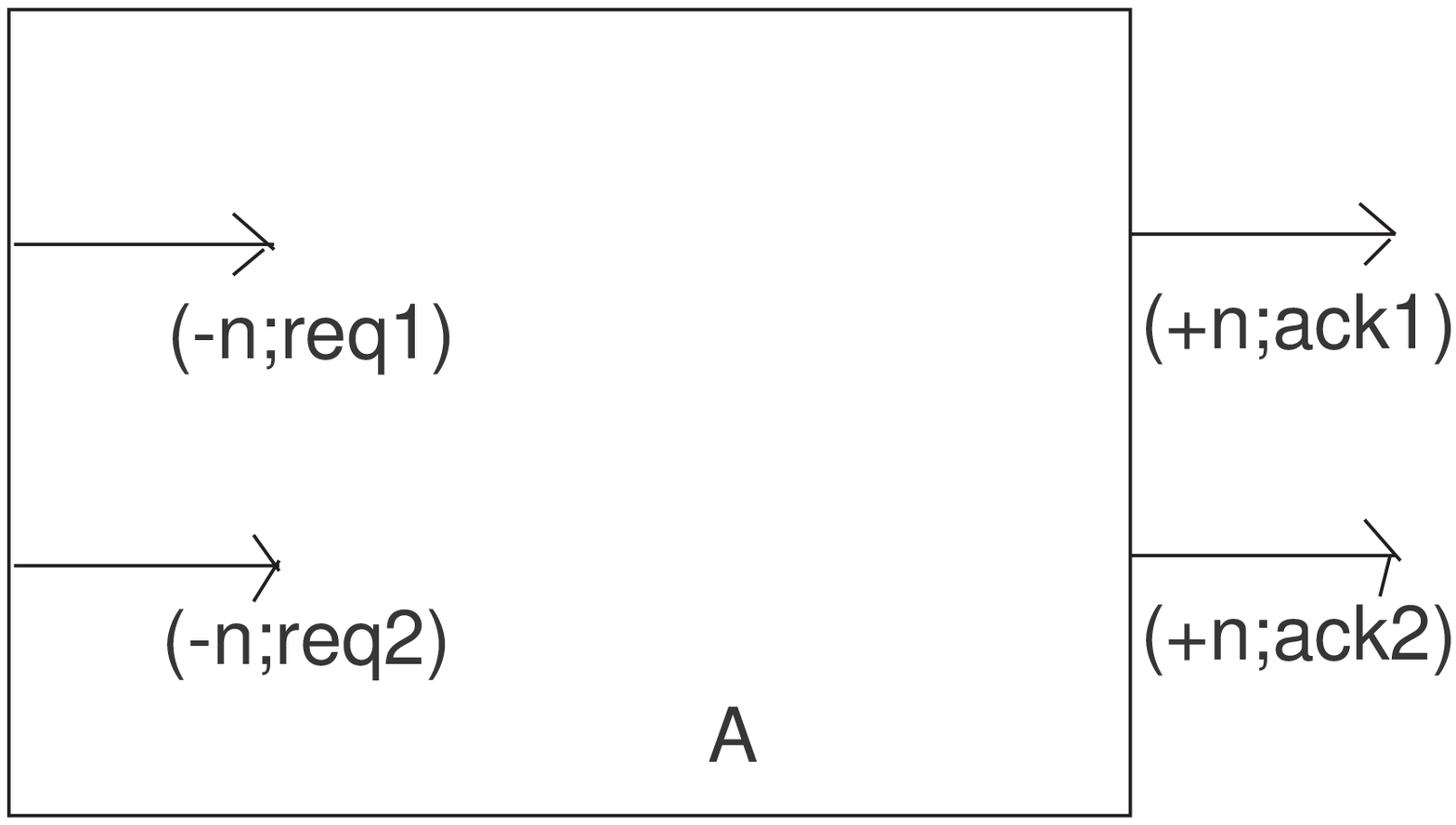}
\end{center}
\caption{Arbiter}
\end{figure}
\\This arbiter's behavior is
(cf. \cite{kgg} \footnote{Up to some minor modifications}) is described by the following {\em complete} set of
rules that contains {\em no redundancies}:
\[\mathcal{S}: \left\{\begin{array}{ll}
{\bf AG}\,[(\neg{\tt req1} \wedge  \neg{\tt req2}) \rightarrow {\bf AX}(\neg{\tt ack1}  \wedge \neg{\tt ack2} )] &
\qquad \rho_1 \\
{\bf AG}\,[({\tt req1} \wedge  \neg{\tt req2}) \rightarrow {\bf AX}{\tt ack1}] & \qquad \rho_2 \\
{\bf AG}\,(\neg{\tt req1} \wedge {\tt req2}) \rightarrow {\bf AX}{\tt ack2}] & \qquad \rho_3 \\
{\bf AG}\,[({\tt req1} \wedge {\tt ack2}) \rightarrow {\bf AX}{\tt ack1}] & \qquad \rho_4 \\
{\bf AG}\,[({\tt req2} \wedge {\tt ack1}) \rightarrow {\bf AX}{\tt ack2}] & \qquad \rho_5 \\
{\bf A}\,[(\neg{\tt req1} \vee \neg{\tt req2}  \vee {\tt ack1} \vee {\tt ack2}){\bf W} ({\tt req1} \wedge
{\tt req2} \wedge \neg{\tt ack1}  \wedge \neg{\tt ack2} \wedge {\bf AX}{\tt ack1} )] & \qquad \rho_6 \\
 ({\tt req1}\wedge {\tt req2} \wedge \neg{\tt ack1}  \wedge \neg{\tt ack2}) \rightarrow {\bf AX}[{\tt ack1}\rightarrow
\\ {\bf A}\,[(\neg{\tt req1} \vee \neg{\tt req2}  \vee {\tt ack1} \vee {\tt ack2}){\bf W}
 ({\tt req1} \wedge {\tt req2} \wedge \neg{\tt ack1}  \wedge \neg{\tt ack2} \wedge {\bf AX}{\tt ack2} )]]& \qquad \rho_7 \\
 ({\tt req1}\wedge {\tt req2} \wedge \neg{\tt ack1}  \wedge \neg{\tt ack2}) \rightarrow {\bf AX}[{\tt ack2} \rightarrow
\\ {\bf A}\,[(\neg{\tt req1} \vee \neg{\tt req2}  \vee {\tt ack1} \vee {\tt ack2}){\bf W}
 ({\tt req1} \wedge {\tt req2} \wedge \neg{\tt ack1}  \wedge \neg{\tt ack2} \wedge {\bf AX}{\tt ack1} )]]& \qquad \rho_8 \\
\end{array} \right.
\]
The computation of the variances for the rules above is summarized in the table bellow:
\begin{figure}[h]
\begin{tabular}{||c|c|c|c|c|c||} \hline \hline
 & \multicolumn{4}{c|}{Message Type} & \\ \cline{2-5}
Rule  & \multicolumn{2}{c|}{Input(s)}  & \multicolumn{2}{c|}{Output(s)} & {$\Delta$} \\ \cline{2-5} %\hline
  & {\tt req1} & {\tt req2} & {\tt ack1} & {\tt ack2} & \\ \hline% & \multicolumn{2}{c|}{{\tt ack1}{\tt ack2} }
 $\rho_1$ & 0 & 0 & 0 & 0 & 0\\ \hline
 $\rho_2$ & 1 & 0 & 1 & 0 & 0\\ \hline
 $\rho_3$ & 0 & 1 & 0 & 1 & 0\\ \hline
 $\rho_4$ & 1 & 0 & 1 & 1 & +1\\ \hline
 $\rho_5$ & 0 & 1 & 1 & 1 & +1\\ \hline
 $\rho_6$ & 1 & 1 & 2 & 1 & +1\\ \hline
 $\rho_7$ & 2 & 2 & 2 & 2 & 0\\ \hline
 $\rho_8$ & 2 & 2 & 2 & 2 & 0\\ \hline \hline
\end{tabular}
\caption{}
\end{figure}
Then $\Delta(A) = \Delta(\rho_1) + ... + \Delta(\rho_8) = +3$. That is: \[\sum_{i \in i(A)}\mu(i) +  \sum_{o \in
o(A)}\mu(o) = +3 \neq 0.\] Thus, in order for ~\ref{eq:THM} to hold, the arbiter also must have a puncture,
responsible resulting from the logical complexity of the block.
\\ Indeed, expressed in the $SMV$ language, the inner structure of the arbiter is given (again cf. \cite{kgg} )
by:
\[ \begin{array}{ll }
   var
\\  \quad req1,req2,ack1,ack2,robin: boolean;
\\ assign
\\  \quad init(ack1) := 0;
\\  \quad init(ack2) := 0;
\\  \quad init(robin) := 0;
\\ next(ack1) := case
\\  \quad !req1 \qquad :0;
\\  \quad !req2 \qquad :1;
\\  \quad !ack1\,\&\,!ack2 \qquad :\,!robin;
\\  \quad 1 \qquad :!ack1;
\\ esac;
\\ next(ack2) := case
\\  \quad !req2  \qquad :0;
\\  \quad !req1  \qquad :1;
\\  \quad !ack1\,\&\,!ack2 \qquad :\,robin;
\\  \quad 1 \qquad :!ack1;
\\ esac;
\\ next(robin) := if\: req1\,\&\,req2\,\&\,!ack1\,\&\,!ack2 \;\; then \;\, !robin
\\ \qquad \qquad  \qquad \quad\ else\: \, robin\:\, endif;
\end{array}
\]
Therefore, a more realistic representation of the arbiter would be  given by  Fig.4:
\begin{figure}[h]
\begin{center}
\includegraphics[scale=0.3]{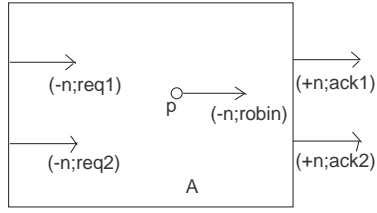}
\end{center}
\caption{An Improved Arbiter}
\end{figure}
Since "{\tt robin}" is asserted iff "{\tt ack1}" or "{\tt ack2}"  are asserted, the signal "{\tt robin}" will
appear three times and, since it is emitted by the puncture towards the arbiter, its sign will be "-". Thus
$\sum_{p \in p(A)}\mu(s) = -3$, as required and, indeed, the fact that the system $\mathcal{S}$ is complete and
contains no redundancies is expressed by the fact that the variance we have found ($\Delta$ = 3) exactly balances
with the contribution of the internal logic due to the "{\tt robin}" puncture.
\\ We further test our method by applying it on the same variations of the main example as considered in \cite{kgg} and
concisely comparing the results.
\\ The first variation is produced by replacing rule $\rho_4$ by $\rho_4'$, thus considering the modified system $\mathcal{S'}$, and also modifying the internal structure of the arbiter by inserting the
new lines bellow: (Here and in the following examples the new/modified rules appear in bold characters.)
\[\mathcal{S'}: \left\{\begin{array}{ll}
{\bf AG}\,[(\neg{\tt req1} \wedge  \neg{\tt req2}) \rightarrow {\bf AX}(\neg{\tt ack1}  \wedge \neg{\tt ack2} )] &
\qquad \rho_1 \\
{\bf AG}\,[({\tt req1} \wedge  \neg{\tt req2}) \rightarrow {\bf AX}{\tt ack1}] & \qquad \rho_2 \\
{\bf AG}\,(\neg{\tt req1} \wedge {\tt req2}) \rightarrow {\bf AX}{\tt ack2}] & \qquad \rho_3 \\
{\bf AG}\,[({\tt req1} \wedge {\tt ack2}) \rightarrow {\bf AX}({\tt ack1 \vee ack2})] & \qquad \huge{\mathbf{\rho_4'}} \\
{\bf AG}\,[({\tt req2} \wedge {\tt ack1}) \rightarrow {\bf AX}{\tt ack2}] & \qquad \rho_5 \\
{\bf A}\,[(\neg{\tt req1} \vee \neg{\tt req2}  \vee {\tt ack1} \vee {\tt ack2}){\bf W} ({\tt req1} \wedge
{\tt req2} \wedge \neg{\tt ack1}  \wedge \neg{\tt ack2} \wedge {\bf AX}{\tt ack1} )] & \qquad \rho_6 \\
 ({\tt req1}\wedge {\tt req2} \wedge \neg{\tt ack1}  \wedge \neg{\tt ack2}) \rightarrow {\bf AX}[{\tt ack1}\rightarrow
\\ {\bf A}\,[(\neg{\tt req1} \vee \neg{\tt req2}  \vee {\tt ack1} \vee {\tt ack2}){\bf W}
 ({\tt req1} \wedge {\tt req2} \wedge \neg{\tt ack1}  \wedge \neg{\tt ack2} \wedge {\bf AX}{\tt ack2} )]]& \qquad \rho_7 \\
 ({\tt req1}\wedge {\tt req2} \wedge \neg{\tt ack1}  \wedge \neg{\tt ack2}) \rightarrow {\bf AX}[{\tt ack2} \rightarrow
\\ {\bf A}\,[(\neg{\tt req1} \vee \neg{\tt req2}  \vee {\tt ack1} \vee {\tt ack2}){\bf W}
 ({\tt req1} \wedge {\tt req2} \wedge \neg{\tt ack1}  \wedge \neg{\tt ack2} \wedge {\bf AX}{\tt ack1} )]]& \qquad \rho_8 \\
\end{array} \right.
\]
\vspace*{0.5cm}
\[ \begin{array}{ll }
   var
\\  \quad req1,req2,ack1,ack2,robin: boolean;
\\ assign
\\  \quad init(ack1) := 0;
\\  \quad init(ack2) := 0;
\\  \quad init(robin) := 0;
\\ next(ack1) := case
\\  \quad !req1 \qquad :0;
\\  \quad !req2 \qquad :1;
\\  \quad !ack1\,\&\,!ack2 \qquad :\,!robin;
\\  \quad \mathbf{robin \,\& \,ack2 \qquad: \{0,1\};}
\\  \quad 1 \qquad :!ack1;
\\ esac;
\\ next(ack2) := case
\\  \quad !req2  \qquad :0;
\\  \quad !req1  \qquad :1;
\\  \quad !ack1\,\&\,!ack2 \qquad :\,robin;
\\  \quad \mathbf{ack2 \qquad :!next(ack1);}
\\  \quad 1 \qquad :!ack1;
\\ esac;
\\ next(robin) := if\: req1\,\&\,req2\,\&\,!ack1\,\&\,!ack2 \;\; then \;\, !robin
\\ \qquad \qquad  \qquad \quad\ else\: \, robin\:\, endif;
\end{array}
\]
These changes produce a positive overall variance $\Delta(A') = 3 > 0$; thus indicating the unbalance introduced
due to redundancy the System of Laws fitting, to the "{\em Unimplemented Transition Evidence}" considered in
\cite{kgg}). The next example relates to the so called "{\em Unimplemented State Evidence}" of \cite{kgg}. It is
produced by introducing an internal auxiliary variable of "input" type , thus augumenting the internal complexity
of the arbiter, in a way that can not be detected and balanced by the rules. In this case the modifications bellow
 indeed generate a  negative $\Delta(A'')$, as expected.
\[ \begin{array}{ll}
   var
\\  \quad \mathbf{req1_{-}temp,req2,ack1,ack2,robin: boolean;}
\\ \mathbf{define \;req1 := req1_{-}temp\,\&\,!(ack1\,\&\,ack2);}
\\ assign
\\  \quad init(ack1) := 0;
\\  \quad init(ack2) := 0;
\\  \quad init(robin) := 0;
\\ next(ack1) := case
\\  \quad !req1 \qquad :0;
\\  \quad !req2 \qquad :1;
\\  \quad !ack1\,\&\,!ack2 \qquad :\,!robin;
\\  \quad \mathbf{ack1 \qquad :\{0,1\};}
\\  \quad 1 \qquad :!ack1;
\\ esac;
\\ next(ack2) := case
\\  \quad !req2  \qquad :0;
\\  \quad !req1  \qquad :1;
\\  \quad !ack1\,\&\,!ack2 \qquad :\,robin;
\\  \quad 1 \qquad :!ack1;
\\ esac;
\\ next(robin) := if\: req1\,\&\,req2\,\&\,!ack1\,\&\,!ack2 \;\; then \;\, !robin
\\ \qquad \qquad  \qquad \quad\ else\: \, robin\:\, endif;
\end{array}
\]
\\ Finally, we consider the modification bellow:
\[ \begin{array}{ll }
   var
\\  \quad \mathbf{req1,req2,req11,req21,ack1,ack2,ack11,ack21,robin: boolean;}
\\ assign
\\  \quad \mathbf{init(ack1) := 0; \;init(ack11) := 0;}
\\  \quad \mathbf{init(ack2) := 0; \;init(ack21) := 0;}
\\  \quad init(robin) := 0;
\\ \mathbf{define}
\\ \mathbf{ack1 := case}
\\  \mathbf{\quad !req11 \qquad :0;}
\\  \mathbf{\quad !req21 \qquad :1;}
\\  \mathbf{\quad !ack11\,\&\,!ack21 \qquad :\,!robin;}
\\  \mathbf{\quad 1 \qquad :!ack11;}
\\ esac;
\\ next(ack2) := case
\\  \mathbf{\quad !req21  \qquad :0;}
\\  \mathbf{\quad !req11  \qquad :1;}
\\  \mathbf{\quad !ack11\,\&\,!ack21 \qquad :\,robin;}
\\  \mathbf{\quad 1 \qquad :!ack11;}
\\ esac;
\\ next(robin) := if\: req1\,\&\,req2\,\&\,!ack1\,\&\,!ack2 \;\; then \;\, !robin
\\ \qquad \qquad  \qquad \quad\ else\: \, robin\:\, endif;
\\ \mathbf{next(req11) := req1;\; next(req21) := req2};
\\ \mathbf{next(ack11) := ack1; \;next(ack21) := ack2};
\end{array}
\]
Since it is basically produced by multiplying the true original signals, the resulting $\Delta(A''')$ is a indeed
a multiple of the original (corresponding to the "{\em Many to One}" (cf. \cite{kgg}).
\begin{remk}
While the existence of punctures is remarked in \cite{kgg} (the so-called "{\em Non-Observable Implementation
Variables}"), the approach described in the mentioned article can  not detect them. This emphasizes one of the
strengths of the method proposed here: it not only detects the above mentioned punctures, but it also estimates
them numerically.
\end{remk}
\section{Global Theory} Since the coverage problem is more crucial in large scale systems, i.e. for units composed of several blocks, it is natural to try to extend the method presented here to such systems as well.
\\ This is possible in the same way that Green's Theorem extends from simply-connected regions to multiply connected regions.
In this manner we obtain the following:
\begin{prop}
 Let $U$ be a unit with bounded environment components $B_j$, each of which corresponds to some block component
 of $U$. For each $B_j$ let $\Delta_i$ denote the difference $\Delta_i = \sum_{i \in i(B_j)}\mu(i) - \sum_{o \in
 o(B_j)}\mu(o)$.
\\ Then $U$ satisfies:
\begin{equation}
\sum_j\mu_U(i) - \sum_j\mu_U(o) + \sum_j\Delta_j = 0
\end{equation}
 where $\mu_U(i),
\mu_U(o)$ are the information measures of $U$ w.r.t. its external environment component. In short, if we denote
$\Delta U = \sum_j\mu_U(i) - \sum_j\mu_U(o)$, then the following holds:
\begin{equation}
 \Delta (U) =  -\sum_j\Delta_j
\end{equation}
\end{prop}
\begin{exmp} The example described in Fig. 5 represents schematically the control unit of a {\em A Bus Interface Unit} (cf.
\cite{gls}\,) and its component blocks.
%\begin{center}
\begin{figure}[h]
\begin{center}
\includegraphics[scale=0.4]{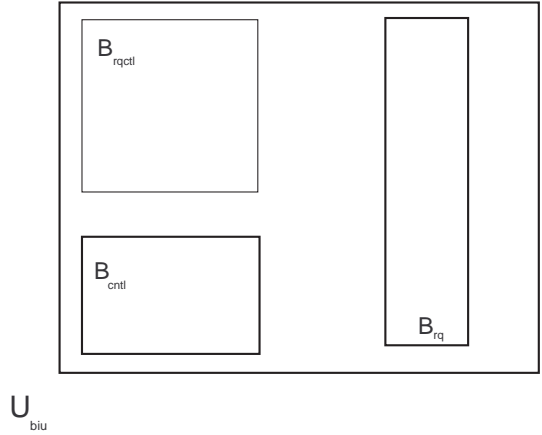}
\end{center}
\caption{The Control unit of a Bus Interface Unit (following [GLS])}
\end{figure}
Then $\Delta (U_{biu}) = -\big(\,\Delta(B_{cntl}) + \Delta(B_{rq}) + \Delta(B_{rqctl}) \big)$
\end{exmp}
Given the technique above, it is evident how to proceed "upward" for larger and larger units: we consider an
integrated circuit $S$ as top level $S = S_0 = L_0$, its composing units as the first level $L_1 = \{S_{1,m}\}$,
their structural subunits as the 2-nd level $L_2 = \{S_{2,n}\}$, and so on,  where, at the "$k$"-th level "$S_k$"
denote the elementary blocks, so eventually we have the following generalization of Theorem~\ref{thm:THM}:
\begin{thm}
\begin{equation} \Delta(S) = \Delta(S_0)  = -\sum \Delta(S_1) = \sum \sum (\Delta(S_2)) = \ldots = (-1)^k\underbrace{\sum
\cdots \sum}_{k\,sums}\Delta(S_k)
\end{equation}
\end{thm}
\begin{exmp} The example presented in Fig. 6 shows  a {\em A Bus Interface Unit} (cf. \cite{gls}\,). The whole
Processor $S$ is designated as level 0\,($L_0$
), the $BIU$ $U_{biu}$ being one of he components of level
1\,($L_1$). The drawing(scheme) also shows the building blocks of $U_{biu}$, which belong to Level 2\,($L_2$).
\begin{figure}[h]
\begin{center}
\includegraphics[scale=0.4]{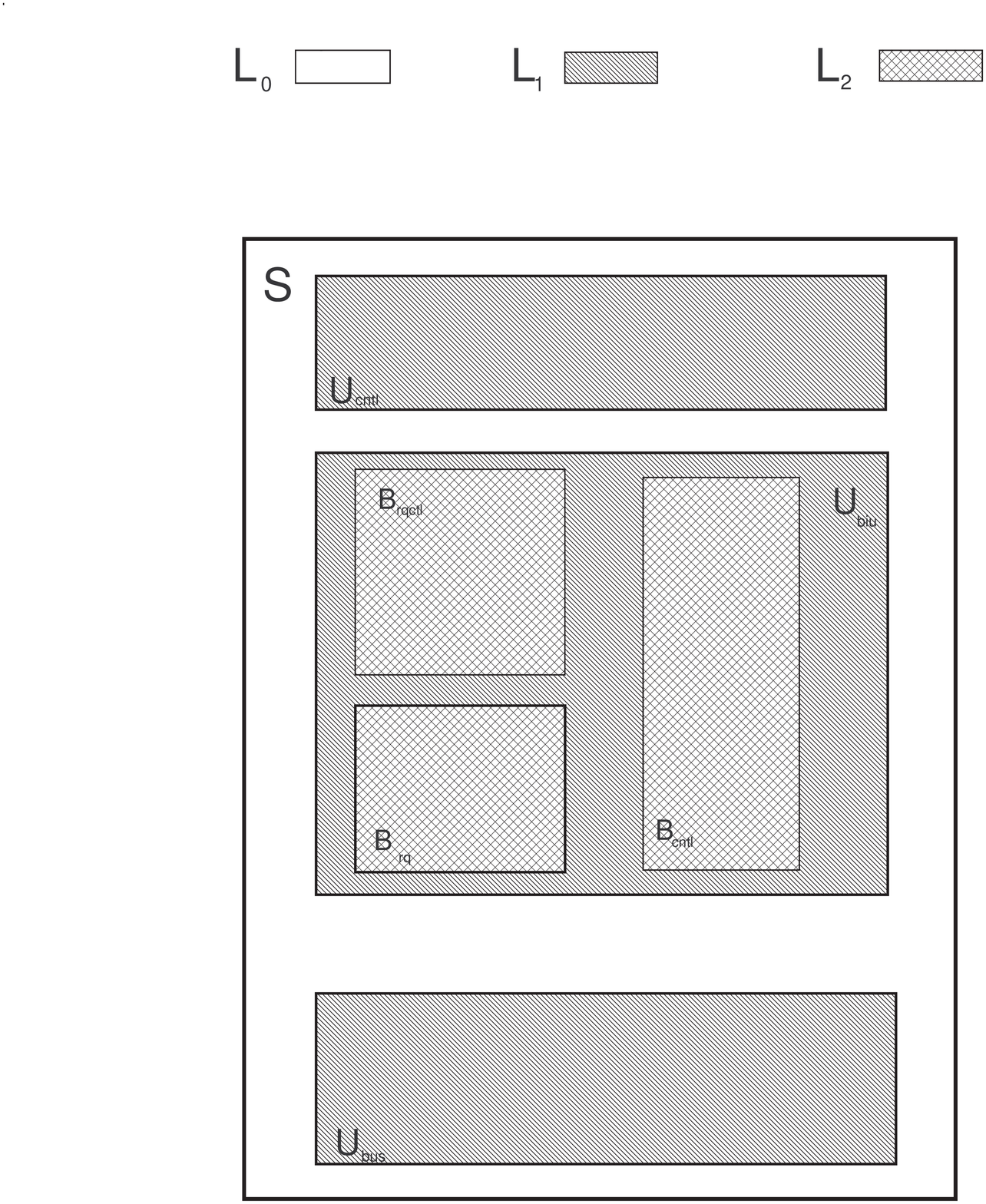}
\end{center}
\caption{A Bus Interface Unit (after [GLS])}
\end{figure}
\end{exmp}
\begin{remk}
Theorem 4.3. gives the verifier the ability to encompass a global estimate viewpoint of the complexity of a large
system, "from top to bottom", as the formula can be readily used at the architectural stage, through the design
faze, down to the verification stage where. At each stage the more complex units are being characterized by having
large pressure contributions. Thus permitting the immediate extension of Model Checking methods to very large
scale systems in a manner which is point-wise precise.
\end{remk}
\section{Future Work} Since punctures, blocks, units, etc., display the same arithmetic behavior, it is only natural to regard each component at any given level as a puncture of the unit of the component containing it and which belongs to next upper level.
Therefore it appears that the appropriate and promising way to study the intrinsic nature of integrated circuits
would be by means of Networks and Graph Theory. Such study is currently in progress.
\section{Appendix}
\begin{thm}{\bf (Green)}
Let $S = int(S) \subseteq \mathbb{R}^2$ be an open set in the plane and let $P, Q: U = int(S) \rightarrow
\mathbb{R}^2$ continuously differentiable functions. Let $\gamma \subset S$ be a piecewise smooth simple, closed
 curve, and let $R = int(\gamma)$ (i.e. $c = \partial{R}$).
\\Then:
\end{thm}
\begin{equation}
\int \hspace{-0.3cm}\int_{\footnotesize R}\hspace{-0.1cm}\Big(\frac{\partial{Q}}{\partial{x}} -
\frac{\partial{P}}{\partial{y}}\Big)dxdy= \oint_{\partial{R}}\hspace{-0.2cm} Pdx + Qdy
\end{equation}
In vectorial notation (6.1) has the following form:
\begin{equation}
\int \hspace{-0.3cm}\int_{\footnotesize R}\hspace{-0.1cm}div\,\overrightarrow{V}dxdy =
\oint_{\partial{R}}\hspace{-0.2cm}\overrightarrow{V}\cdot\overrightarrow{n}ds
\end{equation}
where $\overrightarrow{V} = (Q,-P)$, $div\,\overrightarrow{V} = \frac{\partial{Q}}{\partial{x}} -
\frac{\partial{P}}{\partial{y}}$ is the {\em divergence} of the vector field $\overrightarrow{V}$,
$\overrightarrow{n} = +\,\bf{n}$ is the unit outer normal to $\partial{R}$, and $ds$ represents the length element
of $\partial{R}$.
\\ The classical interpretation of (6.2) above is the following: $\overrightarrow{V}$ represents the {\em flux
density} of an incompressible fluid, then $div \overrightarrow{V}$ measures the amount of mass transported away
from each point per time unit. This quantity differs from zero only then there are sinks and/or sources. Thus
$\int \hspace{-0.3cm}\int_{\footnotesize R} div \overrightarrow{V}dxdy$ measures the amount of fluid escaping from
(respectively entering) the region $R$ through $\partial{R}$. Therefore (6.2) expresses the Mass Conservation Law
for $R$.
% ----------------------------------------------------------------
%\bibliographystyle{amsplain}
%\bibliography{}

\end{document}